# Kinetically accessible 1D magnetic chains of transition-metal chalcogenides and halides on van der Waals surfaces

Canbo Zong[1,2,#], Deping Guo[3,#], Hua Zhu[1,2], Renhong Wang[1,2], Weihan Zhang[1,2], Jiaqi Dai[1,2], Zhongqin Zhang[1,2], Cong Wang[1,2,*], Xianghua Kong[4,*], Fei Pang[1,2], Zhihai Cheng[1,2], Zhong-Yi Lu[1,2], and Wei Ji[1,2,*]

[1]*Beijing Key Laboratory of Optoelectronic Functional Materials & Micro-Nano Devices, and AI-Driven Quantum Materials Research Center, School of Physics, Renmin University of China, Beijing 100872, China*

[2]*Key Laboratory of Quantum State Construction and Manipulation (Ministry of Education), Renmin University of China, Beijing, 100872, China*

[3]*College of Physics and Electronic Engineering, Center for Computational Sciences, Sichuan Normal University, Chengdu, 610101, China*

[4]*College of Physics and Optoelectronic Engineering, Shenzhen University, Shenzhen 518060, China.*

Emails: wcphys@ruc.edu.cn (C.W.), kongxianghuaphys@szu.edu.cn (X.K.), wji@ruc.edu.cn (W.J.).

**Abstract:** One-dimensional (1D) chains offer unique opportunities for nanoelectronics and spintronics, yet their experimental realization remains challenging because 1D motifs are often thermodynamically disfavored relative to higher-dimensional phases. Here we present a high-throughput first-principles exploration of 1D single-atomic transition-metal chalcogenide and halide chains, screening 6,832 candidates constructed from binary combinations of 28 metals and 8 non-metals. To assess kinetic accessibility, we compare the formation energetics of 1D chains with competing two-dimensional polymorphs at the nucleation stage across relevant chemical-potential windows, using nucleation-stage thermodynamic selectivity as a proxy. This workflow identifies 183 kinetically accessible 1D chains. Interpretable machine-learning analysis reveals two simple stability descriptors as key drivers of 1D stabilization. The accessible chains exhibit diverse magnetic configurations with different magnetic characters. We further uncover their pronounced magnetoelastic couplings, exemplified by CrTe with giant magnetostriction reaching 5.93%. Finally, we show that selected metallic ferromagnetic chains retain robust edge magnetism on superconducting substrates, laying the groundwork for proximity-induced topological superconductivity and Majorana zero modes.

## Introduction

Two-dimensional (2D) van der Waals (vdW) materials have rapidly expanded the family of low-dimensional crystals and enabled discovery of diverse emergent phases [1,2], including magnetism [3,4], charge density waves (CDWs) [5,6], and correlation-driven electronic states [7,8]. A defining characteristic of 2D vdW materials is the pronounced anisotropy between in-plane covalent bonding and out-of-plane vdW-type interactions, which renders their properties highly sensitive to layer number [9,10] and stacking geometry [4]. Extending this anisotropy to the one-dimensional (1D) limit goes beyond a simple reduction in dimensionality. Reduced coordination and finite-size confinements substantially reshape structural stability and electronic behavior [11,12]. In 1D vdW chains, every atomic unit is directly exposed to boundaries and interfaces, amplifying the coupling among lattice, charge, and spin degrees of freedom and thereby enables intertwined magnetic and structural responses that are strongly suppressed in higher dimensions [11–13].

Despite extensive theoretical interest, the experimental realization of 1D vdW magnetic chains remains scarce. For a given chemical composition, 1D and 2D motifs inevitably compete during growth. While extended 2D phases are usually thermodynamically favored in the bulk limit, the decisive competition often occurs at the nucleation stage, where finite-size effects, edge energetics, and chemical-potential constraints govern stability [14]. Under such conditions, finite 1D nuclei can exhibit a lower formation free energy than their 2D counterparts within specific chemical-potential windows, biasing early-stage growth toward 1D motifs through kinetic selection. Recent experiments have demonstrated that this nucleation-stage selectivity can be exploited in practice, for example by spatial confinement in carbon nanotubes [15–17] or by chemical-potential control and edge saturation during early growth [18], enabling the formation of single-atomic magnetic chains that would otherwise be precluded by bulk thermodynamics.

However, a predictive and synthesis-aware understanding of which 1D magnetic chains are experimentally accessible is still lacking. Existing high-throughput studies [19–27]. primarily assess thermal stability through bulk formation energies referenced to elemental phases As a result, competition among distinct structural motifs of the same composition, particularly between finite 1D chains and finite 2D flakes at the nucleation stage, has remained largely unexplored [21,28]. Here we address this gap by developing a nucleation-focused first-principles screening framework for 1D single-atomic transition-metal chalcogenide and halide chains. By benchmarking finite 1D motifs against competing 2D polymorphs across non-metal chemical-potential windows relevant to early growth, we identify a set of 1D chains that are plausibly accessible under realistic synthesis conditions. Building on this nucleation-selected set, we establish chemically transparent and data-driven design rules for experimental accessibility. We further show how these accessible chains host low-dimensional magnetism with pronounced magnetoelastic coupling, and potential platforms for proximity-induced topological superconductivity [29–31]. This work provides a strategy for identifying experimentally plausible 1D magnetic materials while revealing their functional opportunities for emergent magnetic and quantum-relevant for properties at interfaces.

**Results and discussion**

Our high-throughput calculation workflow (Fig. 1a) systematically explore the chemical space of binary 1D single-atomic transition-metal chalcogenides and halides. On the basis of nine single-atomic chain prototypes (Fig. 1b) [22,32,33], we constructed 6,832 candidate chains (Step-1 in Fig. 1a) by combining with 28 metals and eight non-metals, while sampling five representative magnetic configurations (Fig. S1) [32]. After all candidates were fully relaxed, we then assessed nucleation-stage thermodynamic selectivity by evaluating chain formation energies across the admissible non-metal chemical potentials and benchmarking against competing 2D polymorphs of the same stoichiometries (ratios of 1:1, 1:2, and 1:3, see Fig. S2) [18]. A chain was deemed kinetically accessible only if it is energetically favored over all competing phases within a finite chemical-potential window (Step-2). Finally, soft-mode instabilities were eliminated by following one of the eight symmetry-breaking distortion modes (Fig. 1c), yielding dynamically stable motifs confirmed by Γ-point phonon analysis (See Fig. S3) [34,35]. This workflow screens 183 potentially accessible 1D chains, spanning 17 ferromagnets (FM), 46 Néel antiferromagnets (AFMs), 18 AFM-AABBs, two AFM-Halfs, and 100 non-magnetic chains.

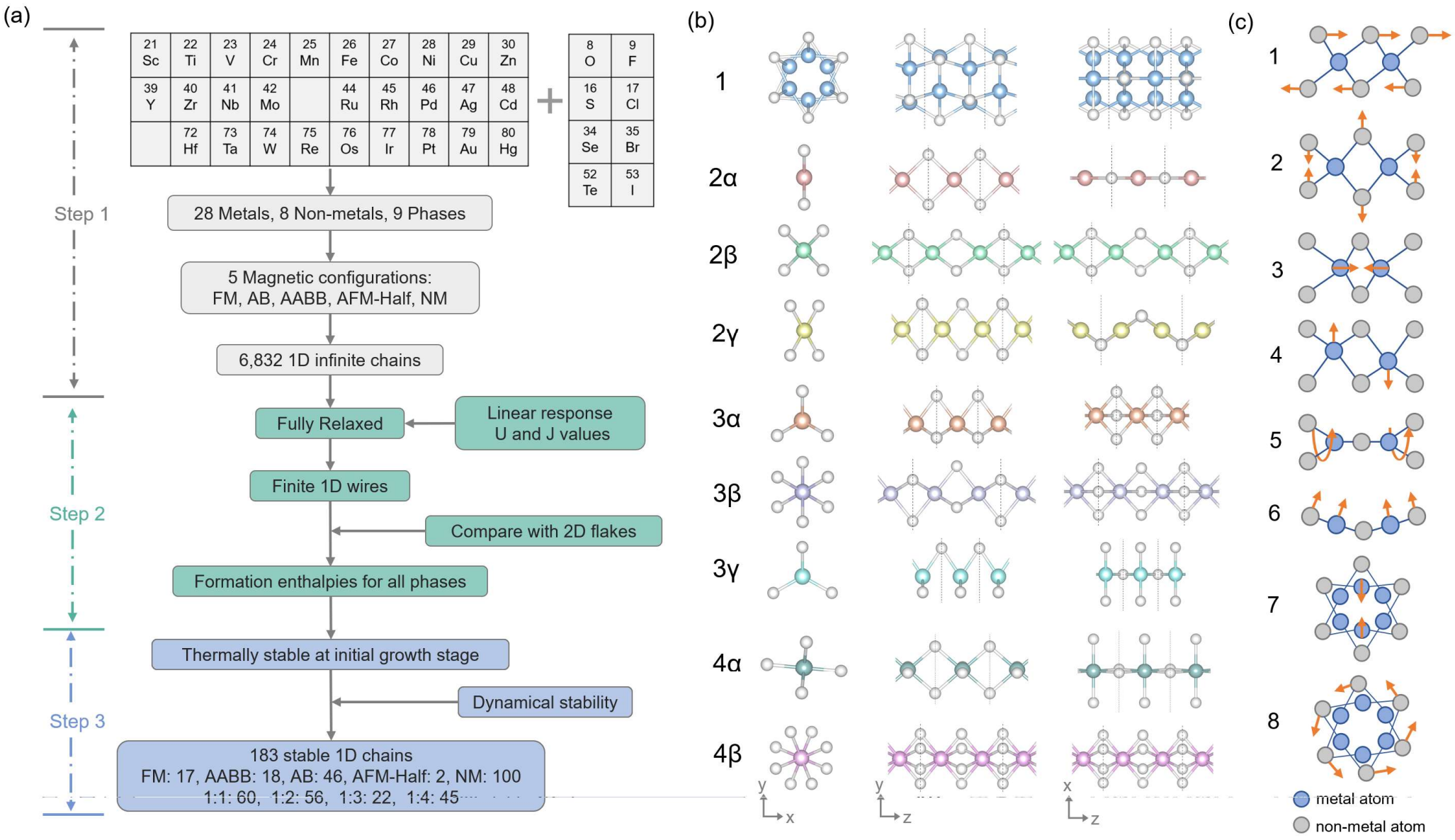

**Fig. 1. High-throughput calculation workflow for discovering kinetically accessible 1D single-atomic magnetic chains.** (a) Schematic of the three steps of our screening procedure. Step 1 (gray) outlines all 28 metal, eight non-metal elements, and five magnetic configurations. Step 2 (green) describes the assessment of kinetic accessibility, where formation enthalpies are computed to construct a stability phase diagram. Step 3 (blue) illustrates the final filtering process with a summary of their favored magnetic configurations. (b) Nine distinct structural polymorphs of 1D chains considered in the screening process. Each row displays the top (left column), perspective (middle), and side (right) views for a given phase, which are labeled from 1 to 4β. (c) Schematic illustrations of the eight structural distortion modes and symmetry-breaking patterns (labeled 1-8). Orange arrows indicate atomic displacement vectors associated with soft phonon modes.

Figure 2 presents a periodic table–style overview of the kinetically accessible single-atomic chains identified in this work. The map organizes the accessible stoichiometries (main numerals) and structural phases (Greek subscripts), together with structural distortion modes (superscript numbers, if appliable), magnetic configurations (symbols, if applicable), and the stability windows in chemical-potential space (background colors). A clear periodic trend emerges: accessible 1D chains are dominated by 3d transition metals (Sc–Zn), which also host the majority of magnetic chains, while 4*d* and 5*d* transition metals yield substantially fewer accessible and magnetic chains. This trend is qualitatively consistent with the enhanced localization of 3d states in 1D, whereas the more extended 4d and 5d orbitals favor delocalization and suppress magnetism.

Beyond the transition-metal periodicity, the accessibility exhibits a strong dependence on anion chemistry. Fluoride chains are kinetically accessible across all transition metals considered, with chlorides showing similarly broad accessibility, whereas accessibility decreases systematically down both the halogen and chalcogen groups, from F to I and from O to Te. Moreover, the accessibility further depends on stoichiometry. Chains with 1:1 composition (MX) are predominantly stabilized under non-metal-poor conditions, while 1:4 chains ($MX_4$) are favored in non-metal-rich environments, indicating that specific stoichiometries can be experimentally selected by tuning the anion chemical potential. Notably, a subset of chains highlighted with orange background in Fig. 2, including MnX (X = S, Se, Te) and $MoX_4$ (X = F, Cl, Br), remains accessible across the entire chemical-potential range, identifying particularly robust candidates for experimental synthesis. These metal-, anion- , and stoichiometry-dependent trends point to common underlying mechanisms associated with nucleation-stage competition between 1D and 2D motifs, whose microscopic origins are elucidated by the machine-learning analysis presented below.

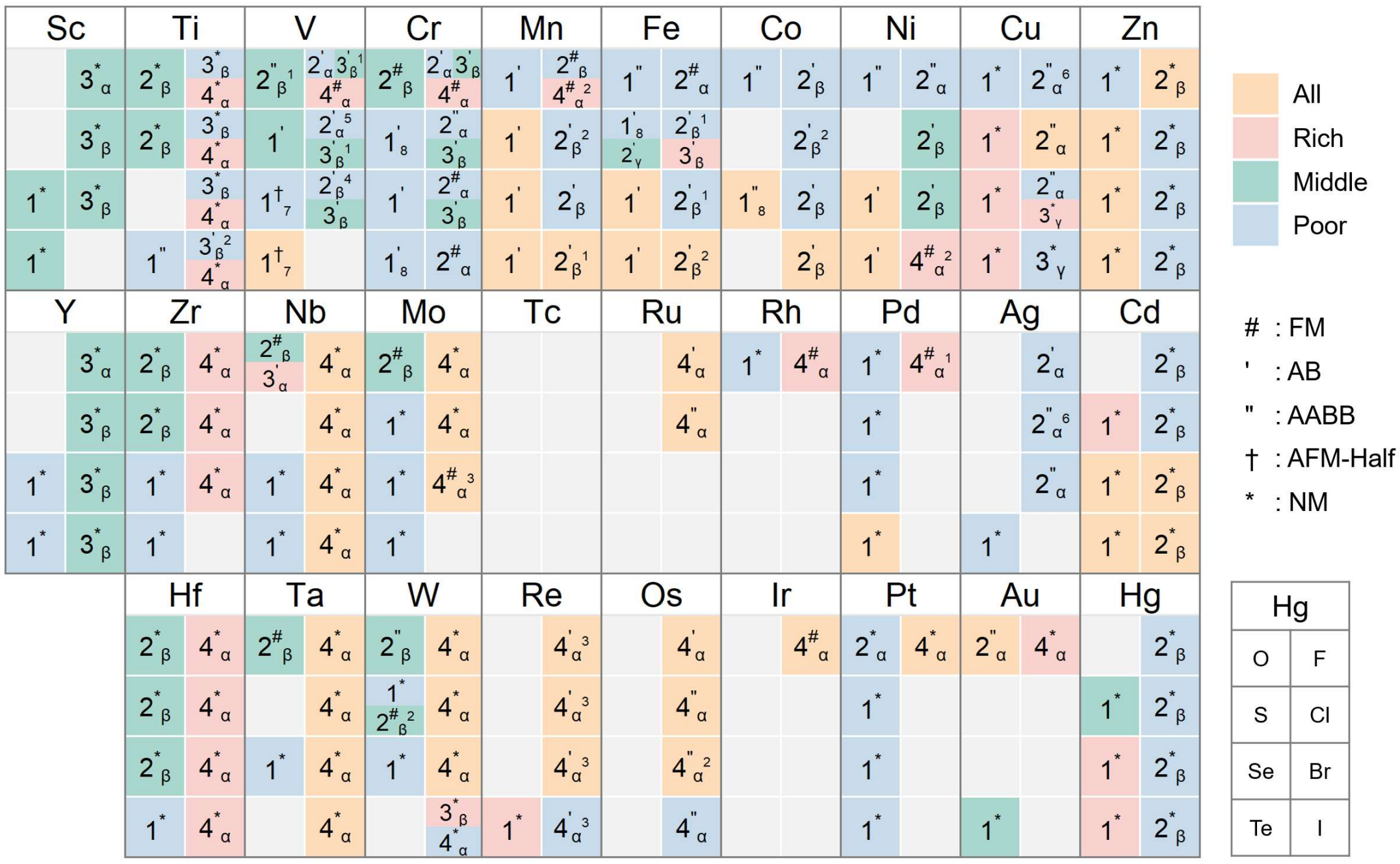

**Fig. 2. A summary map of the predicted kinetically accessible 1D M-X compounds, their properties, and accessibility windows.** The compounds are arranged in a periodic table format, with transition metals (M) as columns and non-metal elements (X) as rows. Each colored cell indicates a predicted accessible compound. The main numeral denotes the chemical ratio (1-4), while the Greek subscript (α, β, γ) indicates the specific structural phase. Small numerical superscripts (for phases 2-4) or subscripts (for phase 1) at the bottom-right of each entry identify the specific structural distortion modes and symmetry-breaking patterns (numbered 1-8, as detailed in Fig. 3). Superscript symbols ’, ”, †, and * indicates ferromagnetism, AB-, AABB-, or Half-type anti-ferromagnetism, and non-magnetism, respectively. Background colors orange, pink, green, and blue represents the kinetic accessibility range with respect to chemical potential of All range, X-Rich, Middle, and X-Poor, respectively.

Using the high-throughput screening results as a labeled dataset, we trained a Random Forest classifier to identify the descriptors controlling nucleation-stage accessibility [36], defined by whether a finite 1D chain is energetically favored over its competing 2D motifs. The feature set included elemental properties of both cation and anion species, including atomic number (AN), density (Dens), atomic radius ($r_{atom}$), electron affinity (EA), ionization energy (IE), and Pauling electronegativity (PE), together with the stoichiometric ratio (Ratio), while the full list is available in Table S1 Feature-importance analysis (Fig. 3a) identifies Ratio as the dominant descriptor (>34%), followed by metal-cation attributes ($Dens_1$, $AN_1$, $r_{atom1}$), and the anion electron affinity ($EA_2$). To resolve the directional influence of these descriptors, we further performed SHapley Additive exPlanations (SHAP) analysis (Fig. 3b) [37,38]. Lower Ratio values contribute positively to accessibility, indicating that compositions with fewer anions per metal preferentially stabilize 1D chains, whereas higher ratios favor higher-dimensional connectivity required to satisfy coordination constraints. Higher $EA_2$ also contributes positively, suggesting that enhanced ability of accumulating electron favors the stabilization of 1D motifs, while larger $GN_1$ (metal group number) contributes negatively, suggesting that a higher valence *d*-electron count disfavor 1D chains and instead stabilize more densely coordinated phases.

Guided by these dominant descriptors, we further sought a compact analytical expression that renders accessibility directly interpretable and predictive (Fig. 3c) [39]. We define a continuous accessibility index $A$ as the fraction of the admissible chemical-potential window over which a 1D chain is thermodynamically favored over competing phases. To ensure the dataset is statistically balanced, we restricted our analysis within period-4 transition-metal chains with the 1:2 stoichiometry, which avoids zero-inflation dominant in broader datasets. Our symbolic regression within the SISSO framework yields a two-descriptor linear model that reproduces the calculated $A$ with high fidelity $R^2$=0.86 and RMSE=0.13 (Fig. 3c): $A = c_1 \underbrace{\left[\left(\frac{1}{EA_2} - \frac{np_2}{IE_1}\right) \cdot \left(\frac{IE_2}{EA_2} - GN_1\right)\right]}_{D_1} + c_2 \underbrace{\left[\frac{AN_1 + np_2 - FN_2^2}{IE_2/EA_1 - FN_2}\right]}_{D_2} + c_0$. The resulting closed-form expression provides a compact and physically interpretable parameterization of the accessibility, serving as a low-dimensional summary of the trends identified above. A detailed analysis of its mathematical structure is provided in Supplementary Note 1.

Taken together, the high-throughput screening, machine-learning analysis, and symbolic regression establishes a unified physical picture in which nucleation-stage competition between one-dimensional and two-dimensional motifs is governed by charge-compensation constraints at undercoordinated boundaries. While finite-size 2D nuclei expose a large fraction of edge units requiring substantial charge redistribution for stabilization, 1D chains are penalized only at two terminal edges and remain energetically favored. This competition is distilled into three governing interpretations where stoichiometry acts as the dominant descriptor because lower anion-to-metal ratios suppress the formation of dense connectivity required for higher-dimensional phases. Additionally, higher anion electron affinity facilitates the metal-anion charge transfer needed to stabilize undercoordinated 1D edges while a lower metal valence electron count avoids the stabilization of more densely coordinated structures. Moreover, the symbolic regression results from the SISSO framework identify the metal-anion electronegativity difference as the primary descriptor of the accessibility. This quantitatively derived framework rationalizes the dependence of 1D accessibility on anion chemistry and stoichiometry observed in our high-throughput screening.

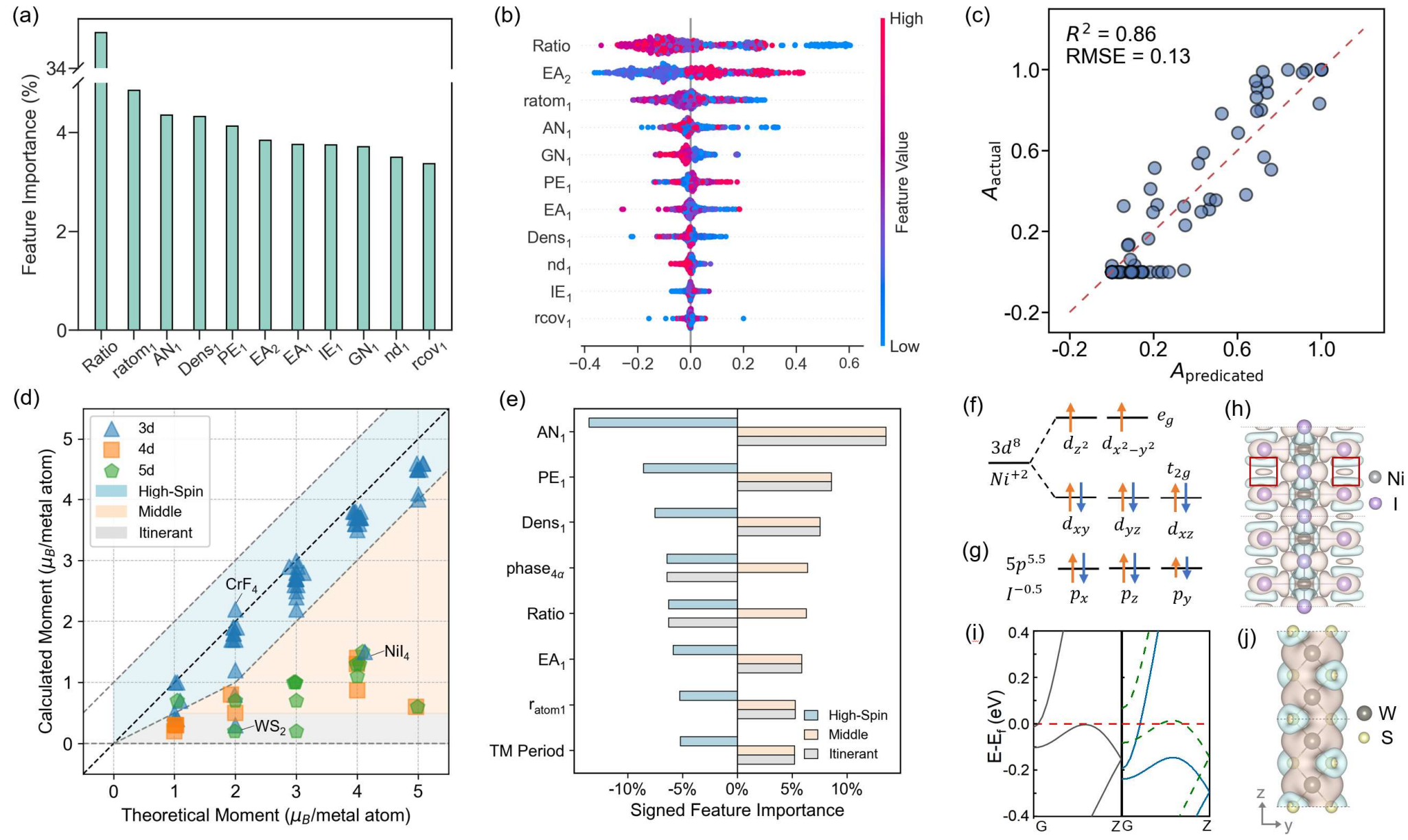

**Fig. 3. Machine learning insights into accessibility, and magnetic classifications of 1D chains.** (a) Feature-importance ranking obtained from a Random Forest classifier trained on the high-throughput screening results. (b) SHapley Additive exPlanations (SHAP) summary plot illustrating the impact of top features on the model output. Each dot represents a sample, with color indicating the feature value (red for high, blue for low). (c) Parity plot comparing the DFT-calculated (actual) and SISSO-predicted Accessibility (*A*). The red dashed line represents the ideal prediction (y=x). (d) Comparison between DFT-calculated local magnetic moments and theoretical values predicted by the ionic model. The background shading distinguishes three regimes based on the deviation from the ionic limit: the High-Spin region (blue), the Middle region (orange), and the Itinerant region (gray). (e) Signed feature importance for magnetic regime classification. (f) Schematic illustration of the $3d^8$ electron filling in $Ni^{+2}$ orbitals within an octahedral crystal field. (g) Schematic illustration of the $5p^{5.5}$ electron filling in $I^{-0.5}$ $5p$ orbitals. (h) Atomic differential charge density of the $NiI_4$ chain. (i) Calculated band structures of the WS2 chain in the non-magnetic (left) and ferromagnetic (right) states. (j) Spin density distribution of the ferromagnetic WS2 chain

We next explore the magnetic properties of the kinetically accessible chains. Three distinct regimes of magnetism were identified by comparing the DFT-calculated local moments with ideal ionic values, as summarized in Fig. 3(d). Halide chains dominated by 3*d* transition-metals (triangles), such as $CrF_4$, cluster near the ionic limit, exhibiting moments that consistent with nominal high-spin configurations and indicative of localized-moment magnetism (blue region) [40]. At the opposite extreme (gray region), chains involving 4*d* or 5*d* metals, such as $WS_2$ and $AuO_2$, display strongly suppressed moments below 0.5 $\mu_B$ per metal atom, reflecting substantial metal-ligand covalency and itinerant electronic character [41,42]. Between these two limits lies a broad intermediate regime (orange region), exemplified by $NiI_4$, where the calculated moments deviate significantly from ideal ionic values, reflecting a competition between localization and itinerancy that invalidates a simple ionic description [43,44].

Feature importance from a Random Forest model (Fig. 3e) identifies atomic number ($AN_1$), Pauling electronegativity ($PE_1$), and the phase-4α structure as primary determinants for the resulting magnetism. Descriptors $AN_1$ and $PE_1$ effectively represent the period (3d, 4d, or 5d) of metal and its ionicity. They are sufficient to distinguish the localized magnetism from others, as the magnetic baseline is intrinsically tied to periodic trends. Moreover, the itinerant and intermediate magnetism are differentiated primarily by the 1:4 stoichiometry (Ratio) and the 4α phase. While $NiI_4$ falls into the intermediate regime of magnetism due to its specific structural constraints, 4α phase chains composed of 4d and 5d metals are mostly categorized into this region with suppresses full electron delocalization. Thus, while atomic identity defines the potential for magnetism, the 4α structure is the critical factor that stabilizes the intermediate-magnetism state across different periods of metals.

Chain $NiI_4$ exemplifies the intermediate-magnetism region. Structural analysis reveals short I–I contacts of 3.1 Å, indicative of appreciable inter-ligand (I-I) wavefunction overlap and charge sharing, as shown in the DCD highlighted by the red boxes in Fig. 3h. Consistent with this picture, orbital-resolved occupancies from projected density of states (Fig. S4) show that Ni adopts a nominal Ni 3$d^8$ configuration (Fig. 3f), corresponding to a +2 oxidation state in an octahedral crystal field rather the formal +4 valence expected from ionic counting. The three $t_{2g}$ orbitals are fully occupied, while the two $e_g$ orbitals ($d_{x2-y2}$ and $d_{z2}$) are half-filled. Integration of the iodine-projected 5$p$ density of states further yields a non-integer occupancy of ~5.5 electrons per I (Fig. 3g; Fig. S4), implying approximately 0.5 ligand holes per iodine that are delocalized over I–I pairs rather than confined to individual sites. Thus, the short I–I contacts and the fractional 5p filling are consistent with the formation of $I_2^-$-like ligand dimers, placing $NiI_4$ in an intermediate regime between localized-moment and itinerant magnetism.

The $WS_2$ chain illustrates the itinerant limit, carrying an approximately 0.25 $\mu_B$/W moment. In the non-spin-polarized calculation, a semi-metallic band structure (Fig. 3i, left) produces relatively flat bands near $E_F$ and a higher density of states (Fig. S5) [45], which yields Stoner parameter, $I \cdot D(E_F) = 9.9$ (where $I = 0.6$ eV and $D(E_F) = 16.5$ states/eV/ W), exceeding the instability criterion of 1 [46]. Allowing spin polarization relieves this instability and leads to an itinerant ferromagnetic state (Fig. 3i, right), with a spin-density extended along the chain rather than localized on atomic sites (Fig. 3j). All these examples demonstrate that the magnetic character of accessible 1D chains is jointly determined by the intrinsic electronic nature of the metal and the specific one-dimensional geometry. While the electronegativity of metals dictates the intrinsic degree of $d$-electron confinement, the specific 1D geometric configuration modulates the transition between localized and itinerant regimes by imposing secondary constraints on electron delocalization.

Beyond their intrinsic magnetic properties, 1D FM chains are of particular interest as building blocks for proximity-induced unconventional superconductivity when interfaced with conventional s-wave superconductors [47,48]. In such heterostructures, superconducting pairing can be induced into exchange-split 1D bands, potentially realizing topological superconducting states [29,49]. A prerequisite for this scenario is the robustness of FM order against substrate-induced perturbations, including epitaxial strain, charge doping, and electronic hybridization [50]. It is, therefore, essential to examine magnetic robustness in explicit heterostructures before discussing proximity-induced topological superconductivity.

We constructed heterostructures by placing kinetically accessible 1D FM chains on four intrinsic vdW superconducting transition-metal dichalcogenides ($NbSe_2$, $TaSe_2$, $NbS_2$, and $TaS_2$). For each substrate, we considered epitaxial alignments along the ***a***-axis and $\sqrt{3}\boldsymbol{a}$ direction, and retained only configurations with lattice mismatch below 3%. Their magnetic outcomes are summarized in Fig. 4a. While some chains undergo substrate-induced magnetic transitions, a substantial fraction preserves FM order on the superconducting substrates (red symbols). Notably, $RhF_4$ and $WS_2$ are metallic on $NbSe_2$ ($TaSe_2$) and $NbS_2$ ($TaS_2$), respectively. As illustrated in Fig 4(b), a finite-length $RhF_4$ chain on bilayer $NbSe_2$ retains uniform spin polarization along the chain, including at its ends, demonstrating that edge moments are not quenched by the superconducting substrate. Moreover, a finite spin polarization is induced in the interfacial Nb layer, indicating appreciable exchange coupling across the interface. These results demonstrate that metallic FM chains can preserve edge-localized spin polarization when supported on vdW superconductors, satisfying a key prerequisite for proximity-induced topological superconductivity.

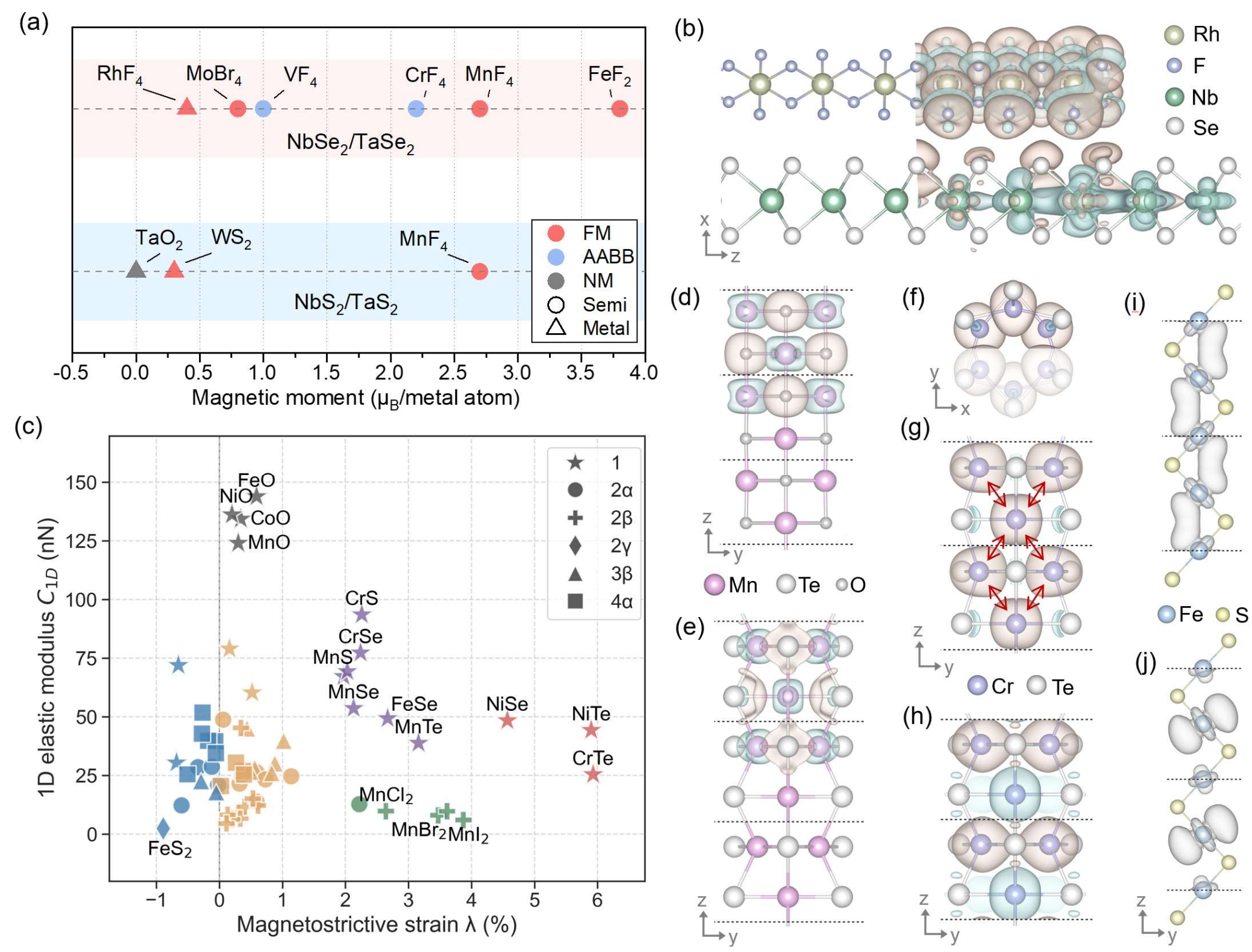

**Fig. 4. Substrate proximity effects and magnetostrictive properties of 1D chains.** (a) Magnetic ground states and local magnetic moments of kinetically accessible ferromagnetic (FM) chains deposited on superconducting transition metal dichalcogenide substrates ($NbSe_2$, $TaSe_2$, $NbS_2$, and $TaS_2$). The shape and color of the points indicate the electronic state (metal/semiconductor) and magnetic order (FM, AABB, or NM) on the substrate, respectively. (b) Spin density of a finite-length FM $RhF_4$ chain on a bilayer $NbSe_2$ substrate. Only half of the finite-length chain is depicted. Furthermore, for clarity, the spin density isosurface (orange for positive, cyan for negative) is mapped for only one-half of this displayed section. (c) The relationship between 1D elastic modulus ($C_{1D}$) and magnetostriction strain ($\lambda$) for antiferromagnetic (AFM) chains. Different colors categorize the materials into distinct mechanical and magnetostrictive regimes: high stiffness (grey), giant $\lambda$ (red), robust magnetostriction (purple), soft with low $\lambda$ (orange), negative $\lambda$ (blue), and soft magnetostriction (green). (d, e) Atomic differential charge density plots for MnO and MnTe, respectively. Orange and cyan regions denote electron accumulation and depletion, respectively. (f) Top-down view of the 1D CrTe chain in the X-Y plane, where the upper half of the chain corresponds to the longitudinal sections displayed in (g) and (h). (g, h) Spin density of the ferromagnetic (FM) and antiferromagnetic (AFM) CrTe chains, respectively. Orange and cyan regions denote spin-up and spin-down densities. (i, j) Partial charge density of 1D $FeS_2$ near the Fermi level. (i) Ferromagnetic (FM) state exhibiting Fe–Fe bonding interactions. (j) Antiferromagnetic (AFM) state lacking specific inter-cation bonding states.

However, the ferromagnetic robustness on superconducting substrates is not universal. As shown by statistics in Fig. 4a, several chains that are FM in isolation undergo substrate-induced magnetic transitions, highlighting the sensitivity of their magnetism to the supporting vdW substrates. Most of these cases are associated with pronounced lattice mismatch, indicating that epitaxial strain plays a dominant role in renormalizing the magnetic order [51]. This observation reflects a strong coupling between magnetism and lattice geometry in a subset of the accessible chains. Chains NiSe, NiTe, and CrTe exhibit exceptionally large magnetostriction strain ($\lambda$) over 4 %, far exceeding typical values (0.015% - 0.04%) reported for magnetic, e.g. Fe-Ge/Cu, nanowires [52]. Here, $\lambda = (L_{\mathrm{FM}}\text{-}L_{\mathrm{AFM}})/L_{\mathrm{AFM}}$, where $L_{\mathrm{FM}}$ and $L_{\mathrm{AFM}}$ represent the equilibrium lattice constants of the FM and AFM states, respectively [53]. To rationalize the magnitude of this magneto-elastic coupling, we analyze the relationship between the 1D elastic modulus ($C_{\mathrm{1D}}$) and the magnetostriction strain ($\lambda$) in Fig. 4(c). Oxides (grey) form a high-stiffness regime with negligible magnetostriction, whereas tellurides (red) and chalcogenides (purple) exhibit giant or robust magnetostriction despite their much lower elastic moduli. The halides (green) represent a soft magnetostriction regime, combining low stiffness with moderate strain response, while the remaining blue and orange groups show negative and minimal $\lambda$, respectively. This systematic classification highlights the diverse magneto-elastic responses of accessible 1D chains and explains the strain-driven magnetic transitions observed on superconducting substrates.

Based on the statistical correlation between magnetostriction and lattice stiffness discussed above, an inverse relationship emerges that enhanced magnetostrictive response is systematically associated with reduced elastic stiffness. This trend follows a chemical progression within the same anion group, where heavier anions, e.g. from O to Te, soften the lattice and amplify magneto-elastic coupling. The microscopic origin of this contrast is illustrated by differential charge-density maps for MnO (Fig. 4d) and MnTe (Fig. 4e). In MnO, charge accumulation is localized around oxygen, reflecting predominantly ionic bonding. Together with a nearly linear Mn-O-Mn geometry, this leads to a rigid structural backbone with a large $C_{\mathrm{1D}}$ (> 100 nN, comparable to the record-high values of ~131–167 nN of diamond nanothreds [54].)

and negligible $\lambda$ (< 1%), which also observed in NiO, CoO, and FeO chains. In contrast, MnTe exhibits more diffuse charge accumulation, consistent with increased covalency. The reduced Mn–Te–Mn bond angle (approximately 123°) introduces geometric flexibility, lowering the elastic modulus and allowing substantial magnetostrictive deformation, thereby accounting for the giant magnetostriction observed in telluride chains.

Chain CrTe offers the highest magnetostrictive response (5.93%) among all considered chains. Comparison of spin-density distributions between the FM (Fig. 4f and 4g) and Neel-AFM (Fig. 4h) states reveals the microscopic origin of this enhancement. In the FM configuration, the majority-spin density on neighboring Cr sites becomes spatially separated (Fig. 4f), arising from exchange-driven avoidance of same-spin charge overlap. The increased short-range cation–cation repulsion favors an expanded lattice (Fig. 4g). However, the AFM configuration allows opposite-spin densities to overlap more efficiently (orange and cyan contour in Fig. 4h), reducing Pauli repulsion and stabilizing a shorter Cr-Cr bond length to 3.11 Å. Consequently, the FM alignment favors an more expanded geometry, characterized by an increased Cr–Cr separation of 3.35 Å and a concomitant widening of the Te-Cr-Te bond angles (from 126.1° to 138.1°), collectively giving rise to the large positive magnetostriction.

An opposite response is found in $FeS_2$, which exhibits the largest negative magnetostriction in our dataset (-1.02%), corresponding to a lattice contraction in the FM state. Partial charge density (PCD) analysis for states near the Fermi level (Fig. 4i–j) shows that, unlike in AFM (Fig. 4j), the FM state hosts bonding states between adjacent Fe atoms along the chain axis (Fig. 4i). These additional bonding interactions provide an attractive interaction that pulls neighboring cations closer, driving a reduction in the Fe–S–Fe bond angle (96.5° → 94.4°) and a shortening of the Fe–Fe distance (3.11 Å → 3.08 Å). These contrasting responses demonstrate that exchange-driven electronic rearrangements can couple to lattice geometry in qualitatively distinct ways, giving rise to both lattice expansion and contraction in magnetically ordered 1D chains.

## Conclusions

We perform high-throughput first-principles calculations on 1D single-atomic transition-metal chalcogenide and halide chains, screening 6,832 candidates built from binary combinations of 28 transition metals and eight non-metals. By benchmarking 1D motifs against competing 2D polymorphs across non-metal chemical-potential windows that capture nucleation-stage thermodynamic selectivity, we identify 183 chains that are preferred within finite growth windows and thus plausible targets for experimental synthesis. Interpretable machine-learning and symbolic-regression analyses distill the dominant descriptors of growth accessibility, with stoichiometry and metal–anion electronegativity contrast emerging as key factors. Nearly half of the accessible chains are magnetic, spanning a continuum from localized to itinerant magnetism with clear chemical trends. These magnets exhibit both positive and negative magnetostriction, including a giant positive value of 5.93% in CrTe and a negative value of −0.89% in $FeS_2$. The contrasting magnetostrictive responses can be understood within a unified picture in which spin-dependent electronic redistribution, constrained by Pauli exclusion and exchange, couples to lattice geometry through bond-length and bond-angle variations in an anion-dependent manner. Gamma-point phonon calculations further verify that identified chains correspond to well-defined local minima without zone-center dynamical instability. Finally, explicit chain-superconductor vdW heterostructure calculations show that several metallic ferromagnetic chains retain the ferromagnetic order and edge spin polarization on superconducting $NbSe_2$/$TaSe_2$ (and $NbS_2$/$TaS_2$) substrates, satisfying key prerequisites for proximity-based topological superconductivity. These results establish a nucleation-focused, physically transparent framework for identifying plausibly accessible 1D materials and for connecting their chemical accessibility with emergent magnetic and quantum-relevant properties.

## Methods

First-principles calculations were performed using the Vienna Ab-initio Simulation Package (VASP) [55,56], employing the projector augmented wave (PAW) method [57]. The exchange-correlation potential was described by the generalized gradient approximation (GGA) in the Perdew-Burke-Ernzerhof (PBE) form [58], with van der Waals (vdW) interactions [59] corrected using Grimme's semi-empirical DFT-D3 scheme. A plane-wave basis set with a kinetic energy cutoff of 700 eV was used for all calculations. To accurately describe the on-site Coulomb interactions of the localized 3d electrons, the DFT+$U$ method was applied to V, Cr, Mn, Fe, Co, and Ni atoms [60]. The effective Hubbard $U$ and exchange $J$ parameters for each compound were determined individually via the linear response method [61], with the specific $U$ and $J$ values for all systems summarized in Table S2.

For the geometric optimization of freestanding 1D chains, we constructed supercells (typically 1×1×2 or 1×1×4) to accommodate various magnetic orderings, ensuring a vacuum layer of at least 15 Å to prevent spurious interactions between periodic images. The Brillouin zone was sampled using a uniform Gamma-centered k-point grid with a linear density of approximately 9.6/$Å^{-1}$. All structures were fully relaxed until the residual forces on each atom dropped below 0.01 eV/Å.

To identify the magnetic ground state, we systematically explored multiple magnetic configurations based on stoichiometry. For 1:1 chains, we considered ferromagnetic (FM), non-magnetic (NM), and three antiferromagnetic states (AFM-ABAB, AFM-AABB, and AFM-Half). For other stoichiometries, the search space included FM, NM, ABAB, and AABB configurations. kinetic accessibility was evaluated by calculating the formation enthalpy (H) relative to competing 2D layered phases, following the formalism established in previous works [18].

To evaluate dynamical stability, phonon frequencies at the Γ-point were calculated using the finite displacement method. For each candidate, atomic displacements of 0.02 Å were applied to construct the Hessian matrix. These calculations were performed within the same supercells used for geometry optimization, maintaining a vacuum layer of at least 15 Å. We employed a Gamma-centered k-point grid with a linear density of approximately 14.3/Å$^{-1}$. To ensure high-precision force evaluation, the electronic energy convergence criterion was set to $10^{-6}$ eV, while all structures were fully pre-relaxed until the residual forces on each atom were less than 0.001 eV/Å.

To determine the 1D elastic modulus ($C_{1D}$) and magnetostriction strain ($\lambda$), we performed strain-dependent energy calculations. The 1D elastic modulus was derived by fitting the total energy ($E$) versus strain ($\varepsilon$) curve to the relation $E(\epsilon) = E_0 + \frac{1}{2} C_{1D} L_0 \epsilon^2$, where $L_0$ is the equilibrium lattice constant and the strain ε ranges from -5 % to +5 %. The resulting $C_{1D}$ is expressed in units of J/m. The magnetostriction strain was calculated as $\lambda = (L_{\mathrm{FM}} - L_{\mathrm{AFM}})/L_{\mathrm{AFM}}$, where $L_{\mathrm{FM}}$ and $L_{\mathrm{AFM}}$ represent the equilibrium lattice constants of the ferromagnetic state and the antiferromagnetic reference state, respectively.

The high-throughput screening workflow, illustrated in Fig. 1, proceeds in three distinct steps. In Step 1 (Grey), we constructed a comprehensive candidate pool by systematically pairing 28 transition metals with 8 non-metal elements across 9 structural prototypes. Considering five potential magnetic configurations for each combination yielded a total of 6,832 infinite 1D chains. In Step 2 (Green), we evaluated the energetic properties of these candidates. Following full structural relaxations using the DFT+U formalism described above, we constructed models for finite 1D wires and competing 2D flakes. The formation enthalpies were then derived by calculating the energy difference between these 1D nuclei and their 2D counterparts, effectively mimicking the energetic competition during synthesis. In Step 3 (Blue), we assessed the kinetic accessibility based on the calculated formation enthalpies. By selecting candidates that are energetically favorable at the initial growth stage, we narrowed the candidate pool down to 183 stable 1D chains, which serve as the basis for the subsequent machine learning and physical property analyses.

To quantitatively identify the key factors governing kinetic accessibility, we employed a Random Forest (RF) classifier model [36] implemented in the scikit-learn library [62]. The model was trained on the high-throughput DFT dataset using a candidate set of 24 initial features constructed from chemical ratio and fundamental elemental properties to predict a binary 'Accessible' or 'Inaccessible' label. The implementation of the model and its parameters, such as handling class imbalance with class_weight='balanced', were based on the scikit-learn library. Statistical robustness was ensured through a repeated stratified 5-fold cross-validation scheme [62] (50 repetitions with different random seeds). The final feature importance scores were determined by averaging the Gini importance values, a standard metric in Random Forests, across all trained models (5 folds × 50 repeats), providing a reliable measure of each feature's contribution to predicting accessibility.

Symbolic Regression via SISSO: To derive an explicit, interpretable functional form connecting atomic features to the computed accessibility (A), we employed the Sure Independence Screening and Sparsifying Operator (SISSO) method [39]. Unlike the global classification task, the symbolic regression was performed on a focused subset of the data comprising transition metal chains from Period 4 with a 1:2 stoichiometry to capture the precise chemical trends within this representative family. The initial feature space $\Phi_0$ consisted of the same primary atomic descriptors used in the RF model. This space was iteratively expanded to a higher-dimensional feature space $\Phi_n$ using a set of unary and binary operators $\{+, -, \times, \div, exp, log\ ,^{-1}\ ,^{2}, \surd\}$. We performed two iterations (n=2) of feature construction. During the screening step, the sure independence screening (SIS) selected the top ranked candidate features based on their correlation with the target A. Subsequently, the sparsifying operator (SO) utilized $L_0$-regularized minimization to identify the optimal 2D descriptor (a linear combination of two complex features) that minimizes the root-mean-square error (RMSE) while maintaining model sparsity.

**Acknowledgements**

We gratefully acknowledge the financial support from the National Key R&D Program of China (Grant No. 2023YFA1406500), the National Natural Science Foundation of China (Grants No. 52461160327 and No. 92477205 ),). Calculations were performed at the Physics Lab of High-Performance Computing (PLHPC) and the Public Computing Cloud (PCC) of Renmin University of China.

# Supplementary information for

# Kinetically accessible 1D magnetic chains of transition-metal chalcogenides and halides on van der Waals surfaces

Canbo Zong[1,2,#], Deping Guo[3,#], Hua Zhu[1,2], Renhong Wang[1,2], Weihan Zhang[1,2], Jiaqi Dai[1,2], Zhongqin Zhang[1,2], Cong Wang[1,2,*], Xianghua Kong[4,*], Fei Pang[1,2], Zhihai Cheng[1,2], Zhong-Yi Lu[1,2], and Wei Ji[1,2,*]

[1]*Beijing Key Laboratory of Optoelectronic Functional Materials & Micro-Nano Devices, and AI-Driven Quantum Materials Research Center, School of Physics, Renmin University of China, Beijing 100872, China*

[2]*Key Laboratory of Quantum State Construction and Manipulation (Ministry of Education), Renmin University of China, Beijing, 100872, China*

[3]*College of Physics and Electronic Engineering, Center for Computational Sciences, Sichuan Normal University, Chengdu, 610101, China*

[4]*College of Physics and Optoelectronic Engineering, Shenzhen University, Shenzhen 518060, China.*

Emails: wcphys@ruc.edu.cn (C.W.), kongxianghuaphys@szu.edu.cn (X.K.), wji@ruc.edu.cn (W.J.).

**Supplementary Note 1: Physical Interpretation of the SISSO Model**

The symbolic regression (SISSO) model distills the complex experimental accessibility into a functional form $A = c_1D_1 + c_2D_2 + c_0$.

$$A = c_1 \underbrace{\left[\left(\frac{1}{EA_2} - \frac{np_2}{IE_1}\right) \cdot \left(\frac{IE_2}{EA_2} - GN_1\right)\right]}_{D_1} + c_2 \underbrace{\left[\frac{AN_1 + np_2 - FN_2^2}{IE_2/EA_1 - FN_2}\right]}_{D_2} + c_0$$

The features analyzed were (where subscripts 1 and 2 denote the metal and non-metal atoms, respectively):

- *AN*: Atomic Number
- *np*: Number of p-valence electrons
- *IE*: Ionization Energy
- *EA*: Electron Affinity
- *GN*: Group Number
- *FN*: Family Number

The learned expression distills experimental accessibility into two primary descriptors, both of which fundamentally encapsulate the electronegativity difference between the metal and non-metal species. The first descriptor, $D_1$, functions as an "electronic matching index" characterizing donor–acceptor compatibility. By coupling the anion's electron affinity ($EA_2$) with the metal's ionization potential ($IE_1$), $D_1$ quantifies the energy-level alignment dictated by the electronegativity mismatch between the two components. The second term, $D_2$, serves as a "penalty term" (where the numerator maintains a negative value across the dataset) modulated by the $IE_2/EA_1$ ratio. This ratio represents the energy cost associated with electronic competition, further refining the accessibility boundary based on the effective electronegativity contrast between the metal and the anion. Together, these terms demonstrate that the formation of 1D motifs is primarily governed by the electronic "mismatch" and resulting charge-transfer potential established by the constituent atoms.

## Supplementary Figures

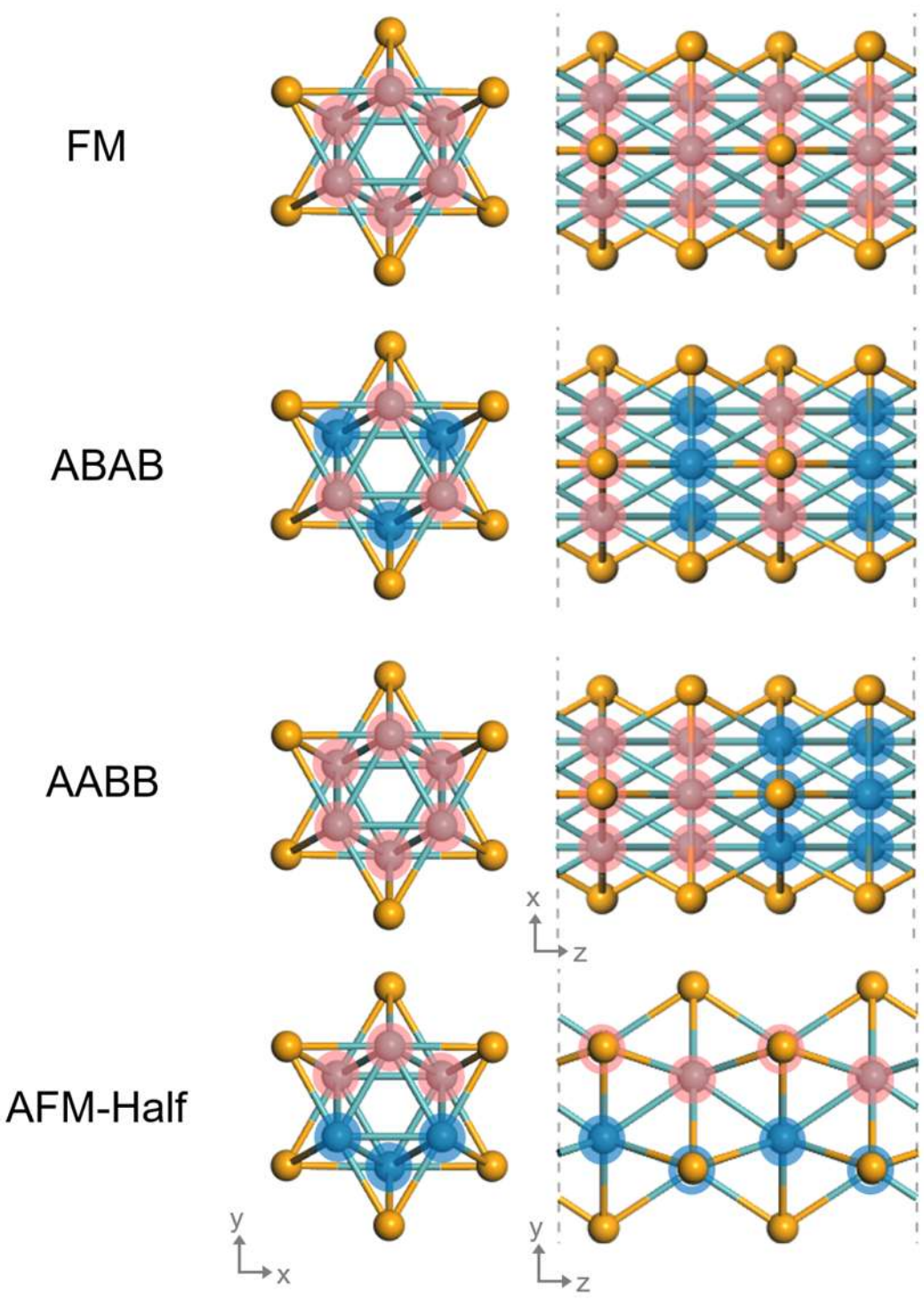

**Figure S1 | Magnetic configurations evaluated for the 1:1 (Phase 1) 1D chains.** Four representative magnetic arrangements are shown: ferromagnetic (FM), and three antiferromagnetic (AFM) patterns labeled as ABAB, AABB, and AFM-Half. Left and right panels display the cross-sectional and longitudinal views of the 1D chains, respectively. The red and blue halos around the metal sites represent opposite spin orientations. These configurations are employed to identify the magnetic ground state and extract the exchange interaction parameters.

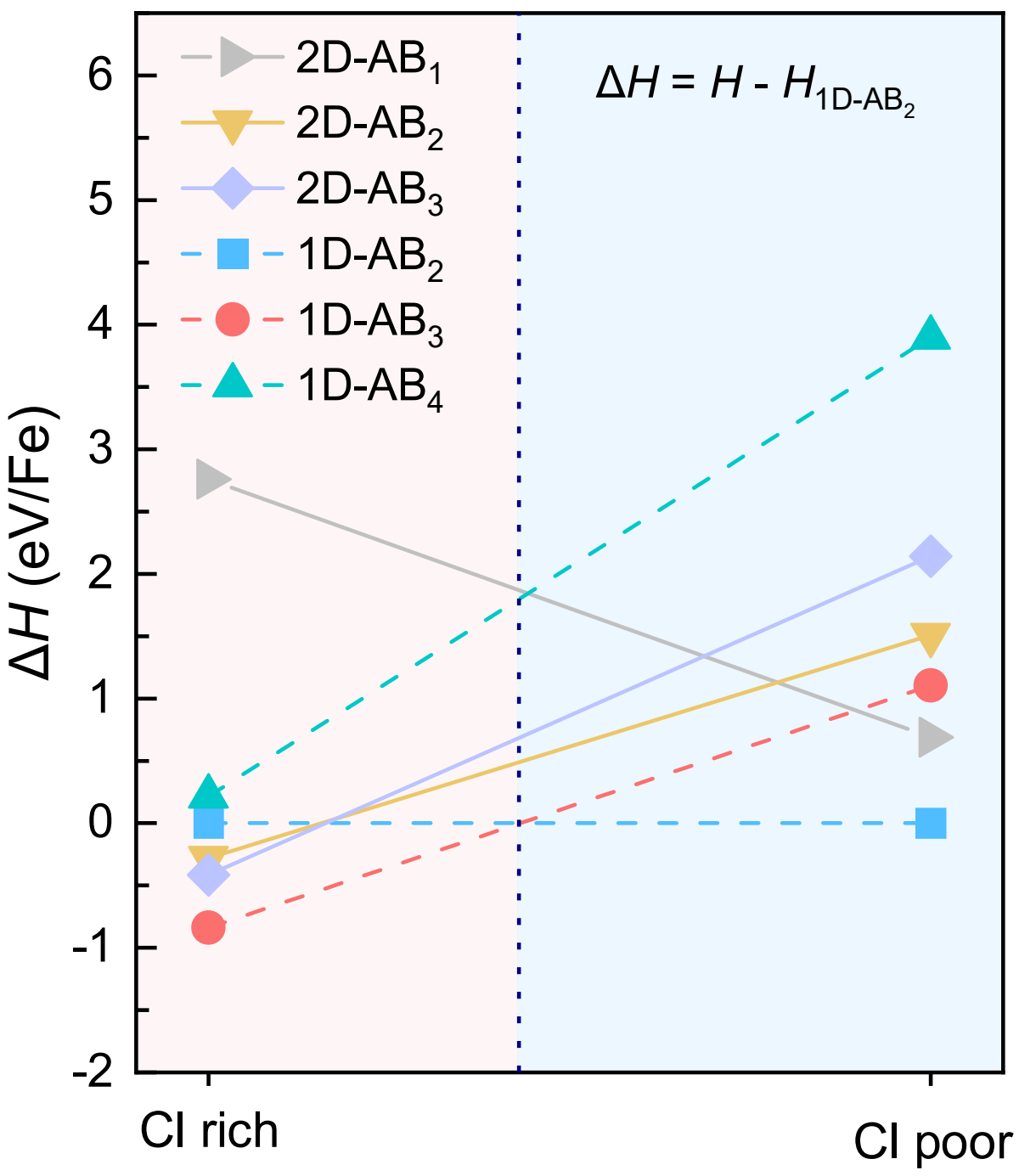

**Figure S2 | Kinetically accessible phase diagram of the Fe-Cl system.** The relative formation enthalpy ($\Delta H$) of various 1D and 2D phases is plotted as a function of the Cl chemical potential, ranging from Cl-rich to Cl-poor conditions. $\Delta H$ is calculated relative to the 1D-$AB_2$ phase ($\Delta H = H - H_{1D-AB_2}$). The shaded regions indicate the stability windows for different 1D motifs: the 1D-$AB_3$ phase (red dashed line) is thermally stable at the initial growth stage under Cl-rich conditions (pink region), while the 1D-$AB_2$ phase (blue dashed line) becomes the most thermally stable phase as the environment transitions toward Cl-poor conditions (blue region). This diagram serves as a representative example for calculating Accessibility ($A$), a continuous descriptor introduced to quantify the experimental feasibility of 1D chains. $A$ is defined as the ratio of the chemical potential range where the 1D chain represents the thermodynamically most stable phase to the total physically allowed chemical potential window. In this specific Fe-Cl system, the calculated Accessibility values for 1D-$AB_3$ and 1D-$AB_2$ are 0.43 and 0.57, respectively.

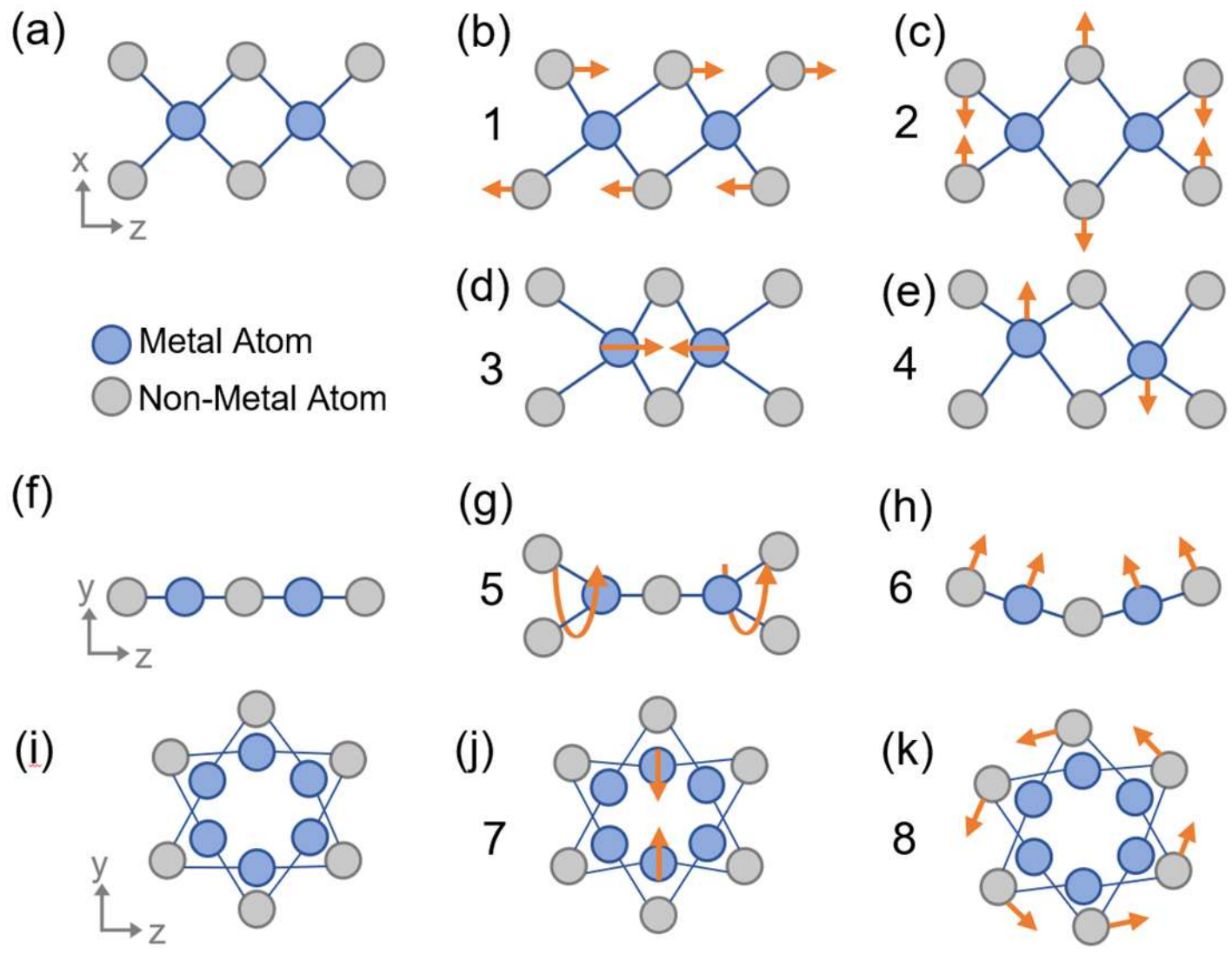

**Figure S3 | Symmetry-breaking displacement modes for generating 1D structural motifs.** (a) The high-symmetry phase-2α prototype structure. (b–e) Primary modes for resolving instabilities in phase-2α: (b) Mode 1 (in-plane non-metal shear), (c) Mode 2 (breathing), (d) Mode 3 (metal dimerization), and (e) Mode 4 (metal shear). (f-h) Collective structural distortions: (f) perspective of phase-2α structure, and (g, h) Modes 5 and 6 represent collective twisting and bending. (i) The high-symmetry phase-1 prototype with C3 rotational symmetry. (j, k) Instability resolution in phase-1: (j) Mode 7 (breaking C3 rotational symmetry) and (k) Mode 8 (breaking mirror symmetry). Blue and gray spheres denote metal and non-metal atoms, respectively; orange arrows indicate the direction of atomic displacements used to generate the library of 183 potentially accessible 1D chains.

Starting from the high-symmetry phase-2α prototype (Fig. S3a), instabilities are typically resolved via in-plane non-metal sliding (shear Mode 1 or breathing Mode 2), metal displacements (dimerization Mode 3 or shear Mode 4), or collective twisting and bending (Modes 5–6). Similarly, instabilities in phase-1 (Fig. S3i) structures are resolved by breaking C3 rotational symmetry (Mode 7) or mirror symmetry (Mode 8). These symmetry-lowered motifs form the structural basis for 183 potentially accessible 1D chains (Step 3 in Fig. 1a).

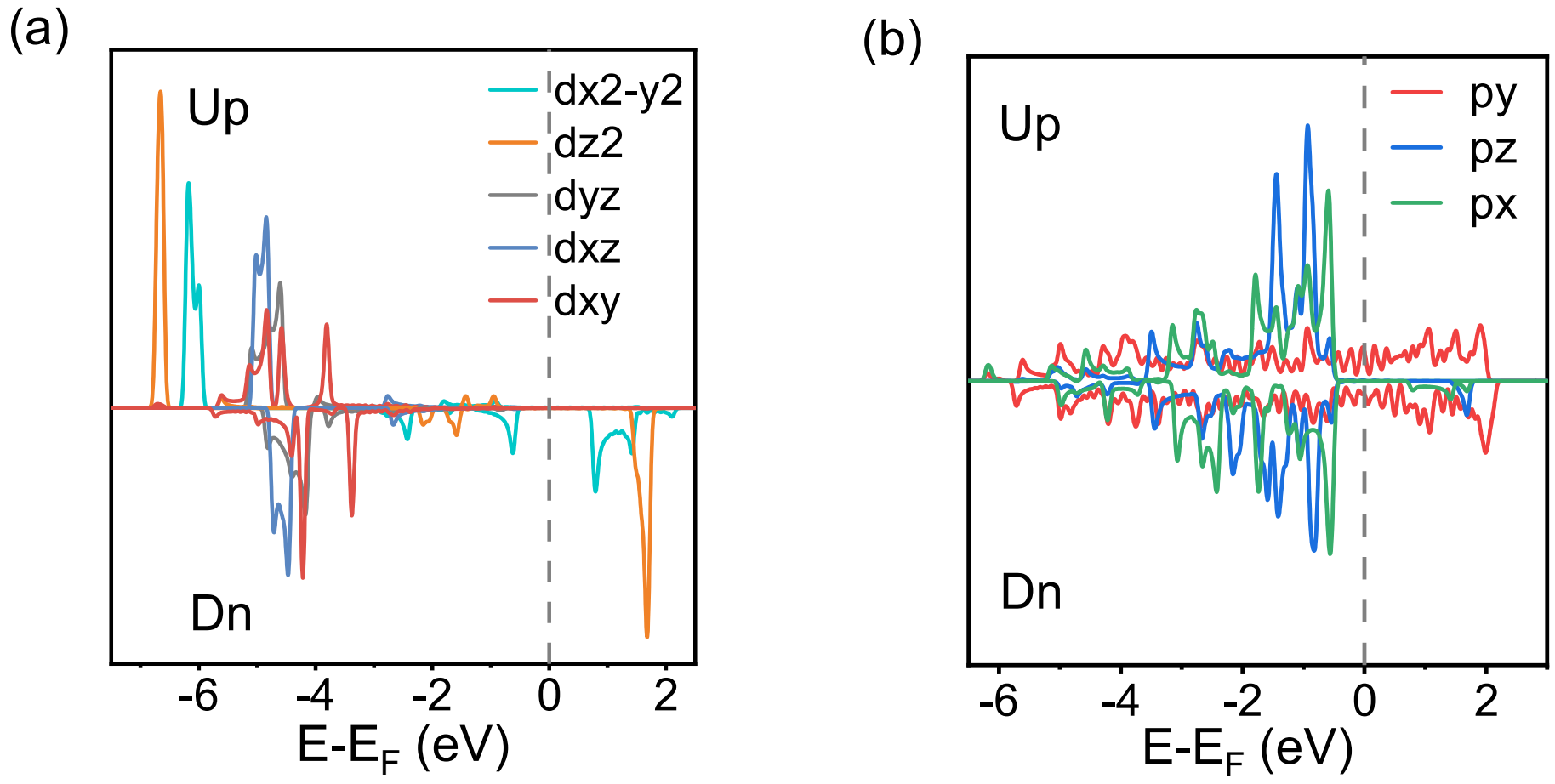

**Figure S4 | Electronic structure of the 1D $NiI_4$ chain.** (a) Projected density of states (PDOS) of the Ni 3*d* orbitals. The Fermi level ($E_F$) is set to zero, indicated by the gray dashed line. The "Up" and "Dn" labels represent the spin-majority and spin-minority channels, respectively. (b) PDOS of the I 5*p* orbitals.

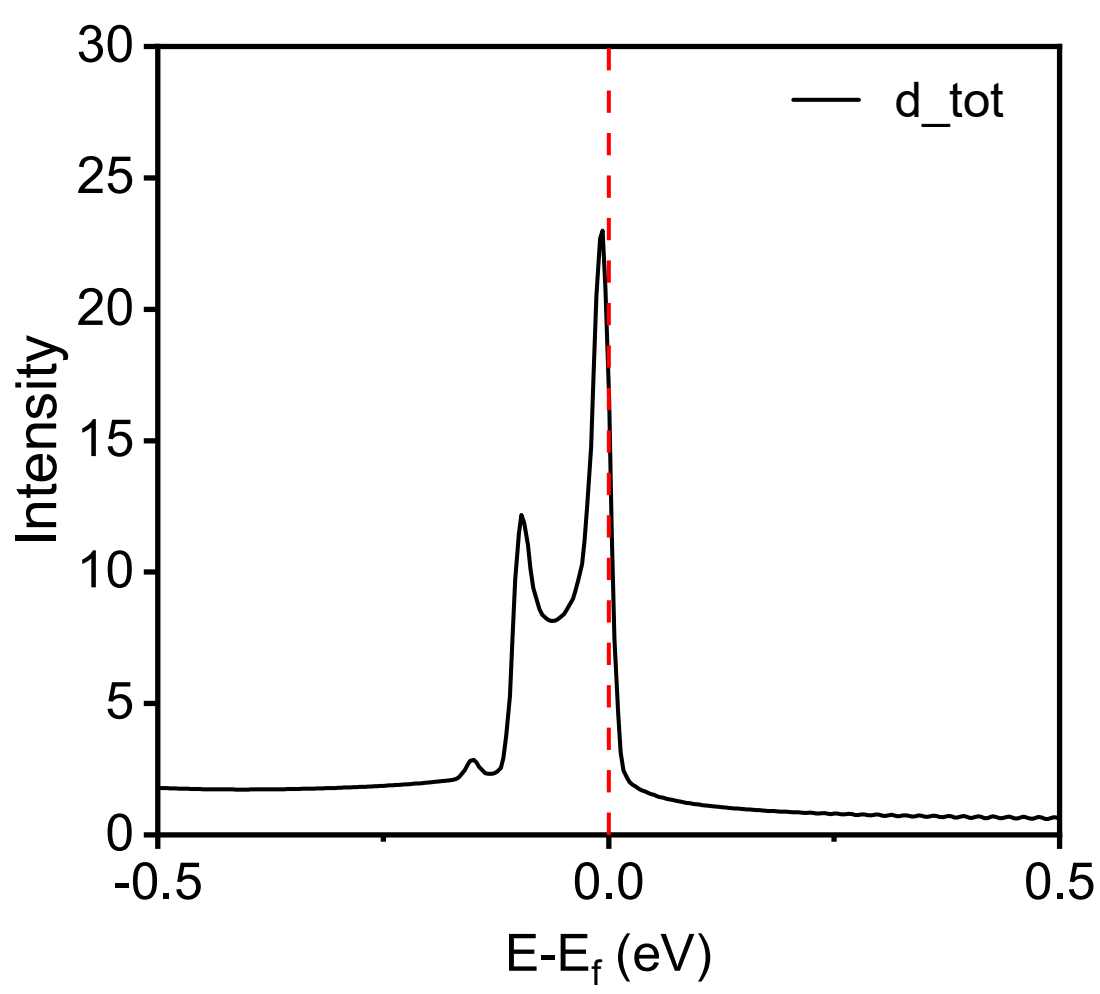

**Figure S5 | Projected density of states (PDOS) of the W d-orbitals in the non-magnetic (NM) $WS_2$ chain.** The Fermi level ($E_F$) is set to zero, indicated by the red dashed line. A sharp, prominent peak in the density of states is observed exactly at $E_F$. This high value of $D(E_F)$ is the physical origin of the large Stoner product ($I \cdot D(E_F) = 9.9$), which triggers the spontaneous exchange splitting and leads to the itinerant ferromagnetic ground state.

**Supplementary Tables**

| **Ratio** | **Dens** | **ratom** | **rcov** | **np** | **nd** |
|---|---|---|---|---|---|
| Chemical ratio | Density | Atomic radius | Covalent radius | Number of p electrons | Number of d electrons |
| **AN** | **GN** | **FN** | **EA** | **PE** | **IE** |
| Atomic number | Group number | Family number | Electron affinity | Pauling electronegativity | Ionization energy |

**Table S1 | Abbreviations and definitions of the atomic and chemical descriptors used in the machine learning models.** These features include fundamental elemental properties and chemical stoichiometry, which serve as the input dataset for both Random Forest classification and SISSO symbolic regression to predict the kinetic accessibility of 1D chains.

|  | V | Cr | Mn | Fe | Co | Ni |
| --- | --- | --- | --- | --- | --- | --- |
| **O** | 4.8 (0) | 5.3 (1.9) | 3.9 (0)* | 5.5 (0.3) | 5.5 (0.9)* | 5.5 (0) |
| **S** | 3.1 (0) | 4.6 (0.6) | 4.1 (0.8) | 5.2 (0) | 7.5 (1.2) | 6.3 (0) |
| **Se** | 2.7 (0) | 4.5 (0.6) | 4.0 (0.7) | 4.9 (0) | 7.7 (1.1) | 6.1 (0) |
| **Te** | 2.2 (0) | 3.0 (0.6) | 3.0 (0.7) | 4.4 (0) | 7.7 (1.3) | 6.2 (0) |
| **F** | 3.7 (0) | 5.9 (0) | 4.9 (1)* | 4.2 (0) | 5.5 (0) | 5.4 (0) |
| **Cl** | 3.1 (0.5) | 5.1 (1) | 4.0 (0) | 4.4 (0) | 6.5 (0) | 5.5 (0) |
| **Br** | 3.9 (0.5) | 5.0 (1) | 3.8 (0) | 4.3 (0) | 6.5 (0) | 5.5 (0) |
| **I** | 3.9 (0.5) | 4.9 (1) | 3.3 (0) | 4.2 (0) | 6.7(0) | 5.8 (0.8) |

**Table S2 | Calculated Hubbard *U* and exchange *J* parameters for 3d transition metal compounds. The values are presented in the format *U* (*J*) with units of electron-volts (eV).** All parameters were determined individually using the linear response method to accurately describe the on-site Coulomb interactions of the localized 3d electrons. Values marked with an asterisk (*) were adopted from literature.